\documentclass[11pt,twoside]{article}

\usepackage{asp2006}
\usepackage{graphicx}
\usepackage{epsfig}
\usepackage{lscape}

\begin{document}
\title{Hot Subdwarfs from SDSS and SPY}
\author{Heiko A. Hirsch and  Uli Heber}
\affil{Sternwarte Bamberg, Sternwartstrasse 7, 96049 Bamberg, Germany}
\author{Simon J. O'Toole}
\affil{Anglo-Australian Observatory, PO Box 296 Epping NSW 1710, Australia}

\begin{abstract}
We present the results of quantitative spectral analysis of subdwarf~O stars
(sdO) obtained from the SDSS database.
By visual inspection, 112 sdO stars could be identified in
a set of spectra of photometrically selected faint blue stars.
Fitting the Balmer, He\,{\sc I} and He\,{\sc II} lines to
state-of-the-art non-LTE model spectra, we derived their effective
temperatures, surface gravities and helium abundances.  We find the
helium-enriched sdO stars to rally in a small intervall of
$T_\mathrm{eff} = 40\,000\,\mathrm{K} \ldots 50\,000\,\mathrm{K}$ and
$\log g = 5.5 \ldots 6.0$, whereas the helium-deficient ones are
showing a wide variance both in temperature and $\log g$.  A puzzling
feature is a significant number of helium rich subdwarfs below the
helium main sequence.  These stars would not be able to burn helium in
their cores and pose a serious problem to subdwarf evolution theories.
We conclude, that the sdB stars are linked with the helium-deficient
sdO stars, i.e.\ the sdBs are predecessors for the helium-deficient
sdOs.
To explain the helium-enriched sdO stars in their entirety, we regard
two scenarios as most promising: The merging of two helium-core white
dwarfs and the late hot flashers.  To investigate this further, we
started an NLTE analysis of carbon abundances.  So far, we find carbon
slightly enriched above solar, but no trends can be seen yet.
\end{abstract}

\section{Introduction}
We distinguish spectroscopically between the cooler sdB stars
($T_\mathrm{eff} < 40\,000\,\mathrm{K}$) showing He\,{\sc I} lines and
the hot sdO stars ($T_\mathrm{eff} > 40\,000\,\mathrm{K}$), with
He\,{\sc II} lines.  Recent work of \citet{stroer07} suggests a strong
correlation of the helium abundance with carbon/nitrogen line
strengths.  A physically more meaningful classification into
`helium-enriched sdO' for stars with supersolar helium abundance and
`helium-deficient sdO' for subsolar abundances is suggested.

While the sdBs form a homogenous group, the sdOs show a wide spread in
temperature, gravity and helium abundance. Only a few stars are
helium-deficient but the helium-enriched sdO stars are represented in
great number.  Canonical theories see an evolutionary link between sdO
and sdB stars, though the mechanism driving the hydrogen-rich sdB
atmosphere into a helium-rich sdO in the course of stellar evolution
is hard to explain.

We are interested in the sdO stars, particulary their evolutionary status, their origins and possible evolutionary links between them and the cooler sdB stars.
Recent work in this field has been carried out by \citet{stroer07} using the high resolution spectra of 46 sdO stars from the ESO SPY survey \citep{napi01}.
We will take advantage of the huge database of the ongoing Sloan Digital Sky Survey which provides spectra of many different objects, among them a number of sdOs.
By comparing and combining our results with the results from \citet{stroer07}, the sample size will be large enough to test rivalling theories for their origin and evolution

\section{Quantitative Analysis of SDSS Spectra}
From the database of the 5th data release (DR5) of SDSS, we selected all point sources within the colour box $(u-g)<0.4$ and $(g-r)<0.1$.
We classified more than 8\,000 spectra by visual inspection, mostly white dwarfs, about 500 sdB stars and 112 sdOs.
After sorting out spectra with spectral features indicative of a cool companion star or of too low signal-to-noise, we carried out a spectral analysis of the remaining 86 of our programme stars.
A model atmosphere fit was conducted in order to obtain the effective temperature, surface gravity and helium abundance.

\subsection{Model Atmosphere Fitting}
\begin{figure}[ht!]
\includegraphics[width=0.49\textwidth]{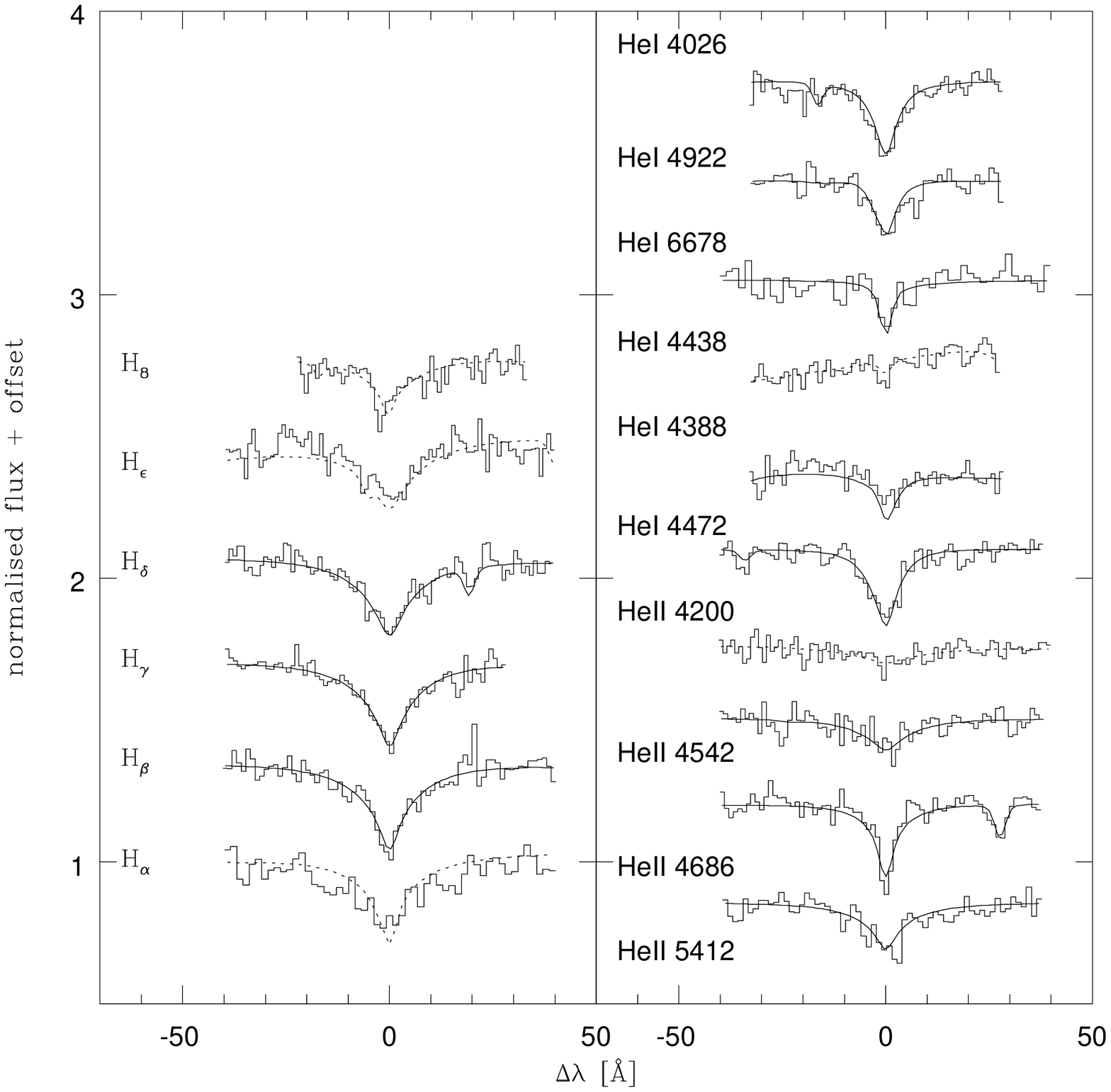}
\includegraphics[width=0.49\textwidth]{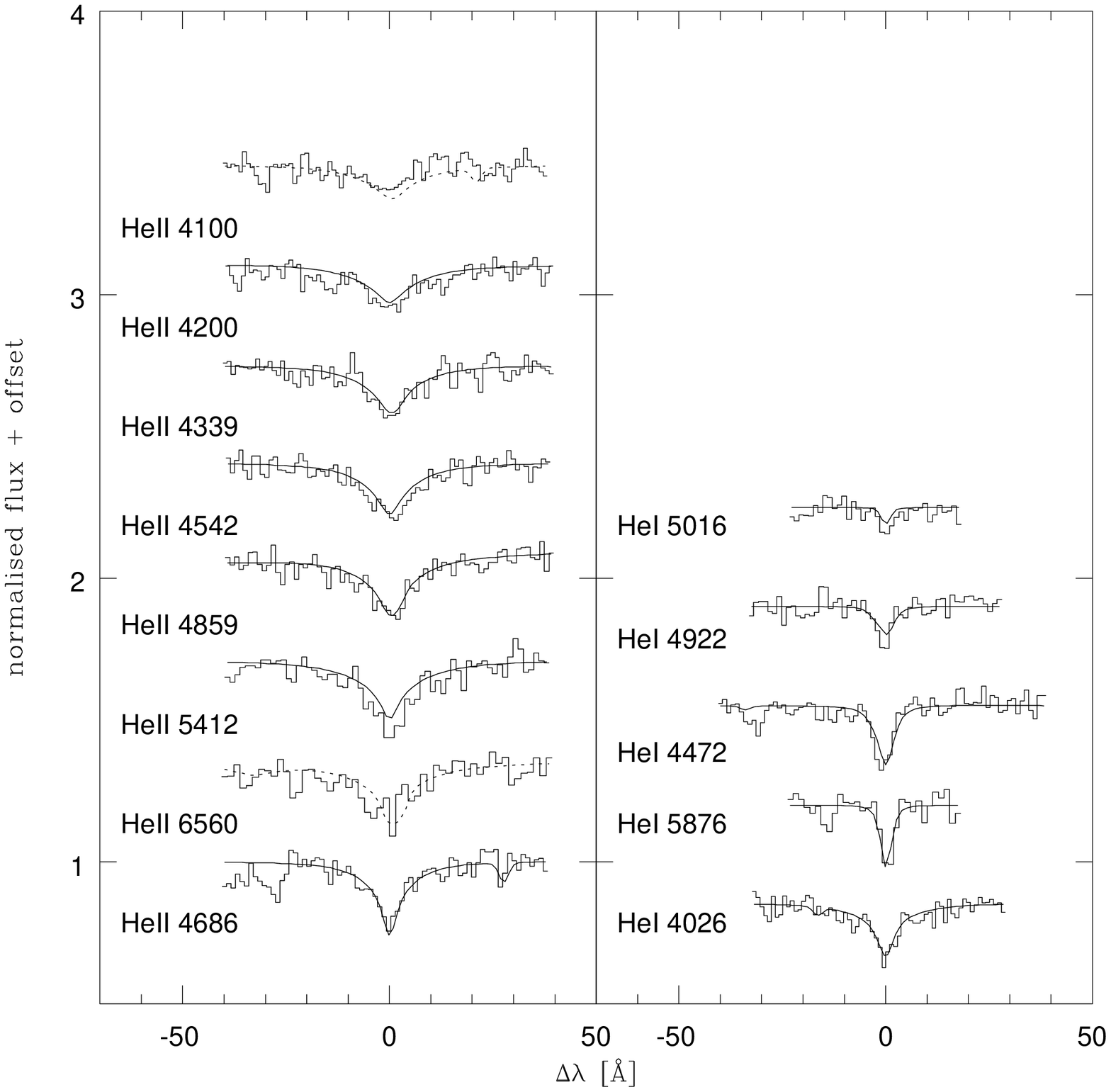}
\caption{Examples of NLTE model fits to sdO spectra from SDSS.  {\it
Left hand side:} a helium-deficient sdO. {\it Right hand side:} a
helium-enriched sdO.  The observed spectra are displayed as
histograms.  Lines that are not fitted but only displayed for
inconsistency checks are plotted as dashed lines.}
\label{img:bspfit}
\end{figure}

To determine $T_\mathrm{eff}$, $\log g$ and $\log
(\frac{N_\mathrm{He}}{N_\mathrm{H}})$ of the stars, we fitted
synthetic model spectra to the observed ones, using a
$\chi^2$-minimization procedure \citep{napi99} to derive all three
parameters as well as the radial velocity simultaneously.  The
synthetic non-LTE models were constructed using the PRO2 code
\citep{werner99} with a new temperature correction technique
\citep{dreizler03} and include partially line blanketing.  It includes
H and He atoms only.  The model grid ranges in temperature from
30\,000\,K to 100\,000\,K and from 4.8\,dex to 6.4\,dex for $\log g$
with $\log (\frac{N_\mathrm{He}}{N_\mathrm{H}}) = -4 \ldots +3$.  No
extrapolation beyond the model grid was allowed.  Due to the
inconsistencies reported for the H$_\alpha$-line of some sdB stars,
which might be caused by a stellar wind \citep{heber03wind}, this line
was never fitted to the models, but it was always plotted to search
for peculiarities.  H$_\beta$ to H$_\gamma$ and He\,{\sc
II}\,4859\,{\AA} and 4339\,{\AA} were always fitted, but we found
He\,{\sc II}\,4100\,{\AA} often too weak to provide a reasonable fit.
The ionization equilibrium of He\,{\sc II} and He\,{\sc I} is very
sensible to temperature, therefore we always included He\,{\sc
I}\,4472\,{\AA}, even in its absence.

\subsection{Considering the Errors}
The fit program provided us with mere statistical errors, they range
from $\Delta T_\mathrm{eff} \approx 150 \ldots 3\,000\,\mathrm{K}$,
depending on the signal--to--noise ratio.  Fortunately, six stars were
observed twice (see Table.~\ref{tab:error}), which we can use to
estimate the quality of our fits.  The differences in parameters
obtained from spectrum A and spectrum B lie within the given
statistical errors, which we therefore conclude to be robust.
\begin{table}[ht!]
\centering
\caption{Two exposures had been taken from six of our programme stars. We use this opportunity to check the reliability of the errors.}
\label{tab:error}
\smallskip
{\small
\begin{tabular}{l r@{} l c c }
\tableline
\noalign{\smallskip}
SDSS designation & $T_\mathrm{eff}$ & & $\log g$ & $\log (\frac{N_\mathrm{He}}{N_\mathrm{H}})$\\
&[K] &&[$\mathrm{cm\,s^{-2}}$]\\
\noalign{\smallskip}
\tableline
\noalign{\smallskip}
SDSSJ 00:51:07.01+00:42:32.5 & 38234 & $\pm$330 & 5.73$\pm$0.06 & $-$1.07\\
& 38207 & $\pm$305 & 5.72$\pm$0.05 & $-$0.94\\
SDSSJ 11:14:38.57$-$00:40:24.1 & 57714 & $\pm$3268 & 5.56$\pm$0.19 & $-$0.94\\
& 55730 & $\pm$2292 & 5.56$\pm$0.15 & $-$0.86\\
SDSSJ 16:37:02.78$-$01:13:51.7 & 45547 & $\pm$338 & 5.77$\pm$0.08 & +3\\
& 45860 & $\pm$353 & 5.70$\pm$0.08 & +3\\
SDSSJ 17:00:45.67+60:43:08.4 & 47404 & $\pm$1870 & 5.92$\pm$0.16 & +2.26\\
& 49537 & $\pm$2911 & 6.09$\pm$0.17 & +2.17\\
SDSSJ 23:35:41.47+00:02:19.4 & 69906 & $\pm$2234 & 5.45$\pm$0.08 & +1.10\\
& 72870 & $\pm$1341 & 5.44$\pm$0.11 & +1.18\\
SDSSJ 23:39:13.99+13:42:14.2 & 48592 & $\pm$1540 & 5.82$\pm$0.15 & +0.92\\
& 48914 & $\pm$3233 & 5.66$\pm$0.18 & +1.19\\
\noalign{\smallskip}
\tableline
\noalign{\smallskip}
\end{tabular}
}
\end{table}

\subsection{Results and Comparison to the SPY Sample}
\begin{figure}[ht!]
\includegraphics[width=0.49\textwidth,bb=18 144 592 688]{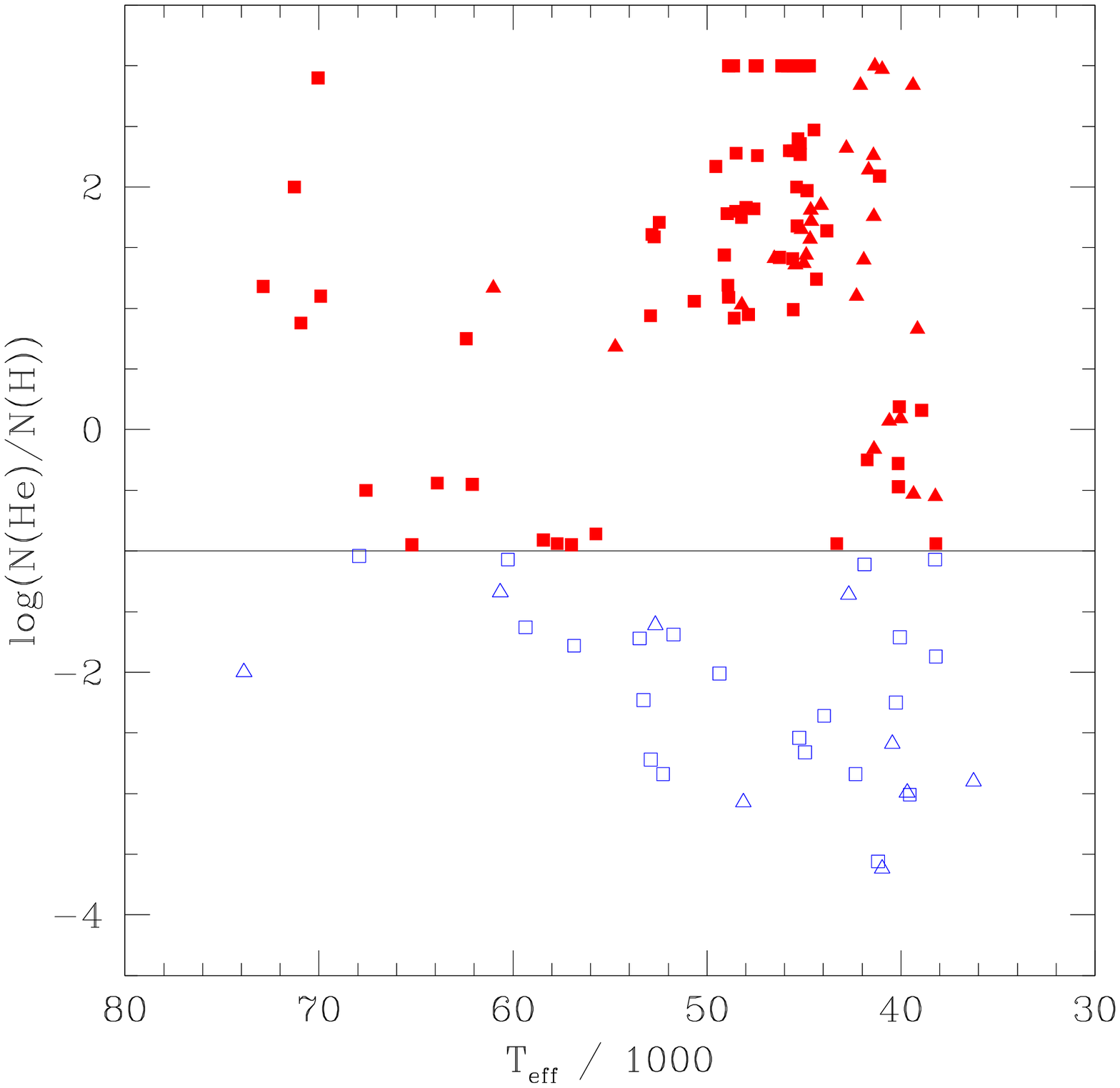}
\includegraphics[width=0.49\textwidth,bb=18 144 592 688]{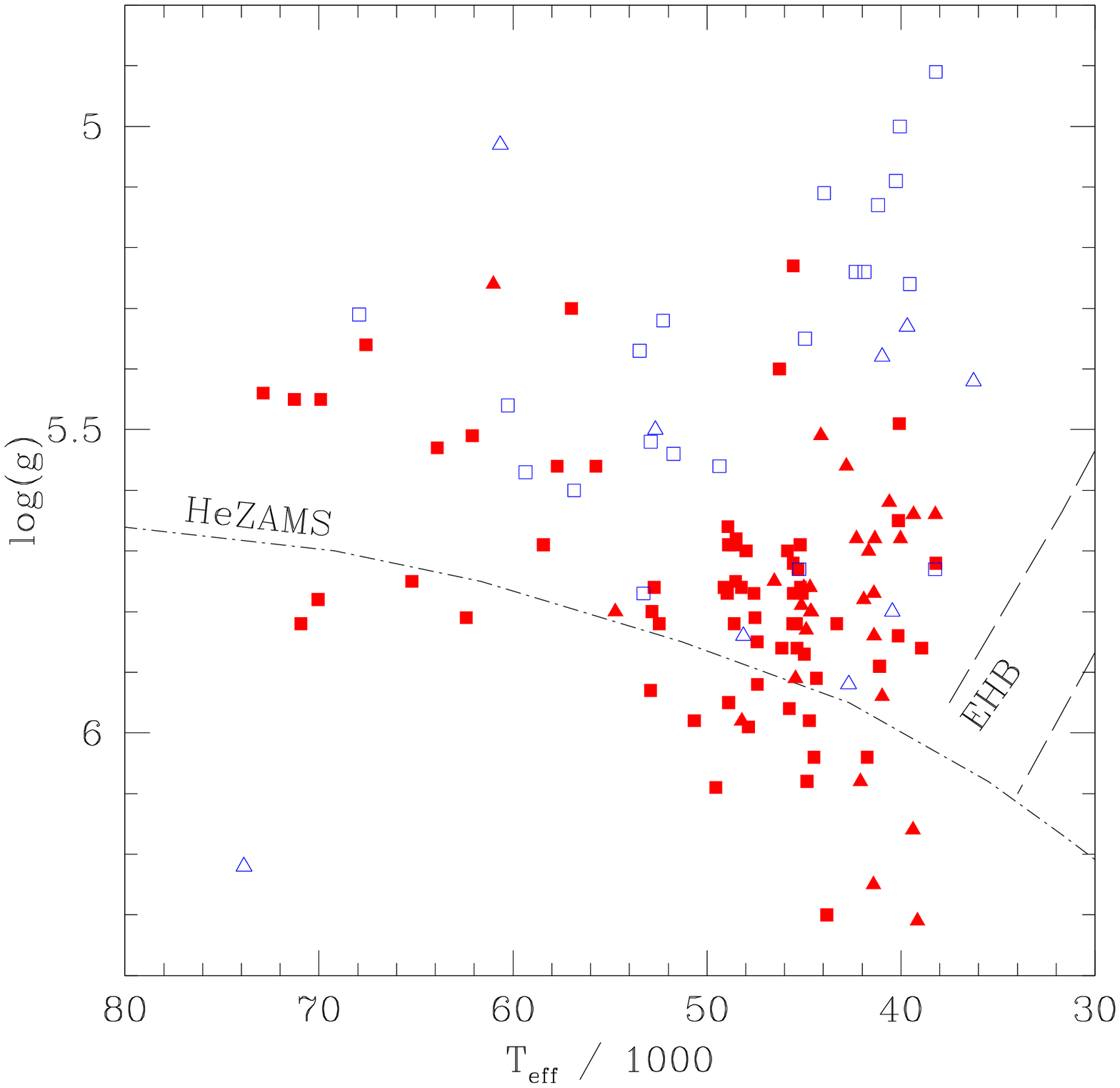}
\caption{{\it Left:} $T_\mathrm{eff}$--$\log
(\frac{N_\mathrm{He}}{N_\mathrm{H}})$ diagram. The solid horizontal
line represents the solar helium abundance.  {\it Right:}
$T_\mathrm{eff}$--$\log g$ diagram. Also given is the helium main
sequence (HeZAMS) as well as the EHB, defined by the location at zero
age and terminal age.  Filled symbols represent helium-enriched stars,
while the open symbols are helium-deficient ones.  Squares are stars
from SDSS, triangles are stars from SPY.}
\label{img:hrd}
\end{figure}

We now compare our SDSS sample to the 46 stars of the SPY sample.  We
find that the ratio of helium-enriched stars to helium-deficient ones
is higher in SDSS than in SPY (68:18 vs.\ 33:13).  In addition, the SDSS
sample contains more very hot stars ($T_\mathrm{eff} >
60\,000\,\mathrm{K}$) than the SPY sample.  Both findings are probably
due to selection effects, e.g. the limiting magnitude of the surveys,
which is considerably fainter for SDSS than for SPY.  Besides that,
the distributions of the SDSS and the SPY sample in the
$T_\mathrm{eff}$--$\log g$ plane are very similar, as can be seen from
Fig.~\ref{img:hrd}.  We will therefore merge the samples for further
discussion.

None of the programme stars is located on the EHB, meaning that either
they are post-EHB stars, most likely evolved from the sdB stars, or
they do not have any link to the EHB at all and are just evolving
through this region in the HR diagram.  The helium-enriched sdOs are
spread over $T_\mathrm{eff} = 40\,000\,\mathrm{K} \ldots
74\,000\,\mathrm{K}$ and $\log g = 5.2 \ldots 6.3$, but a
concentration of them within the small temperature interval
$T_\mathrm{eff} = 40\,000\,\mathrm{K} \ldots 50\,000\,\mathrm{K}$
becomes apparent, only a few stars are found to lie at hotter
temperatures.  The hydrogen-rich sdOs on the other hand shun this
region.

Note that both the SDSS and the SPY sample contains stars lying below the helium main sequence, all but one are helium-enriched.
These stars cannot sustain stable helium burning in their cores and therefore it is difficult to explain their origin.

\section{Stellar Evolution Tests Using the Combined Samples}
Two different aproaches to hot subdwarf star evolution are presented
in the literature: \emph{(i)} binary star evolution and \emph{(ii)}
single star evolution.  Both are trying to explain the high mass loss
the stars have to suffer from before reaching the EHB.  In the context
of binary star scenarios, this is achieved by mass exchange in
sufficiently close systems, whereas single star scenarios have to
invoke physical mechanisms within the star itself, like unusually
strong stellar wind, mixing, etc.

\subsection{Binary Star Evolution}

\citet{han02,han03} did an extensive binary population synthesis and
found three formation channels relevant for the sdB/sdO population:
stable Roche lobe overflow, common envelope ejection and a merging of
two helium-core white dwarfs.  While the first two scenarios result in
relatively cool stars preferentially lying between 30\,000\,K and
40\,000\,K, our sdO stars are very unlikely to have followed the latter
formation channel as they are mostly hotter.

For sdO stars, especially the helium-enriched ones, we regard the
formation via a merging of two helium WDs as most promising.  Short
period binary WDs will lose orbital energy through gravitational
waves.  With shrinking separation, the less massive object will
eventually be disrupted and accreted onto its companion, leading to
helium ignition.  \citet{saio2000} argue, that this merger product
will result in a helium buring subdwarf showing an atmosphere enriched
in CNO-processed matter.  This scenario therefore can explain these
extremely helium-enriched sdOs showing strong nitrogen lines in their
atmospheres.  However, \citet{gour06} find that, under the assumption
of total angular momentum conservation, He+He WD merger do rotate
faster than breakup velocity.  A mechanism that enables the star to
get rid of its angular momentum still has to be found.

However, there is another possible origin for hot subdwarf stars.  The
sdB HD~188112 is reported to have a mass of $0.23\,M_\odot$, based on
a Hipparcos parallax measurement, too low to sustain core helium
burning \citep{heber03}.  Such a star may have formed if mass transfer
occured on the first red giant branch (RGB) and before the onset of
core helium burning.

Apropriate tracks have been calculated by \citet{driebe98}, who
evolved a $1\,M_\mathrm{\odot}$ main sequence star as one component of
a binary system through the RGB stage without ignition of helium to a
helium star and eventually to a helium core white dwarf, under the
assumption of mass transfer to the companion.  The evolution of such
stars with masses between $0.414\,M_\mathrm{\odot}$ and
$0.300\,M_\mathrm{\odot}$ would lead straight through the observed sdO
population, down to below the HeZAMS.  However, how compelling these
findings may be, they suffer from a fundamental discrepancy:
\citet{napi04} find that the fraction of radial velocity variable
stars among the helium-enriched sdOs is 4\,\% at most.  Also, the
predicted helium abundance is much lower than for the helium-enriched
sdOs suitable for this scenario and within the relevant mass range, no
enrichment of the atmosphere with processed matter from the interior
is predicted.

\subsection{Single Star Evolution}

\begin{figure}[ht!]
\centering
\includegraphics[width=0.49\textwidth,bb=18 144 592 688]{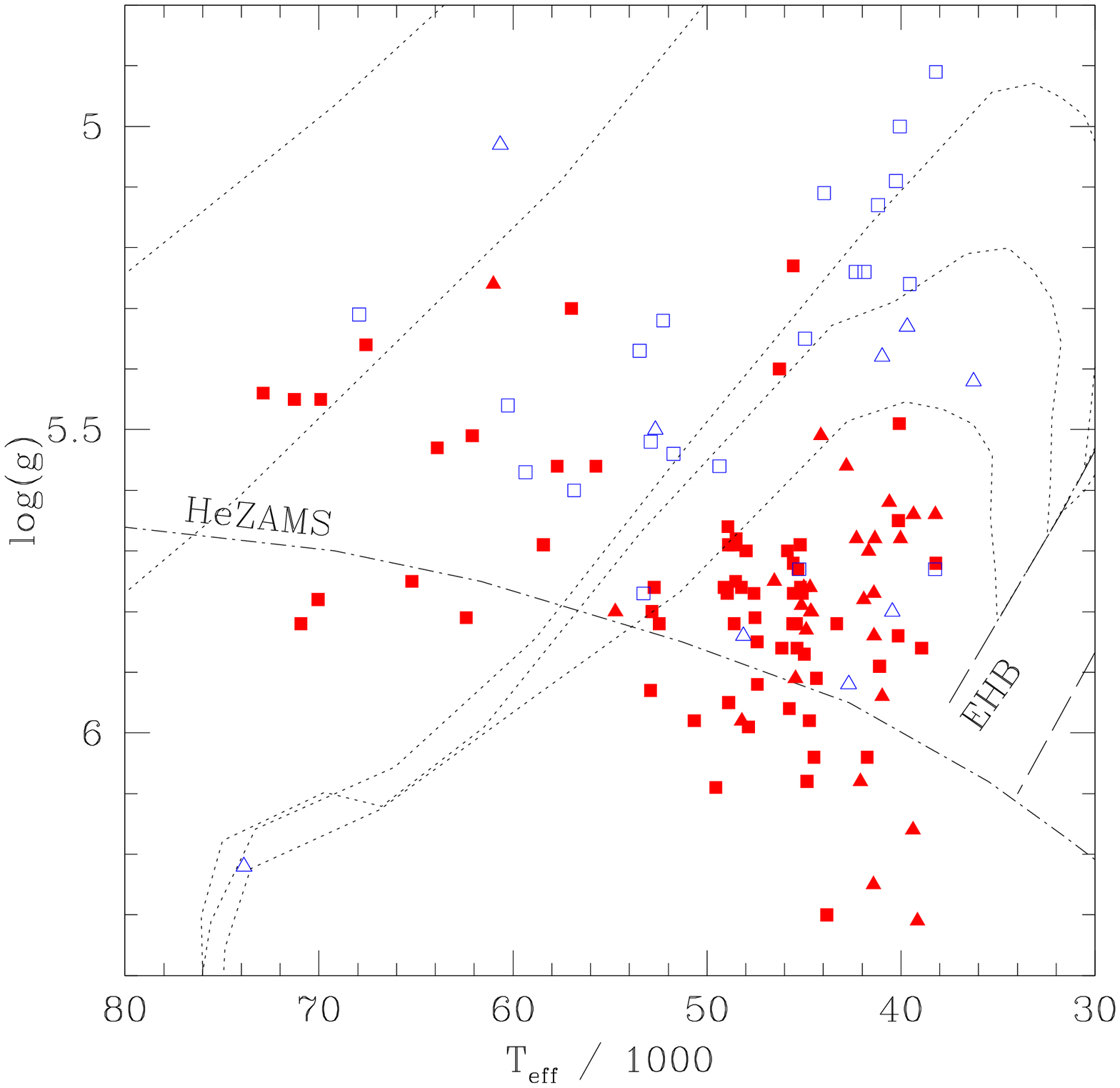}
\includegraphics[width=0.49\textwidth,bb=18 144 592 688]{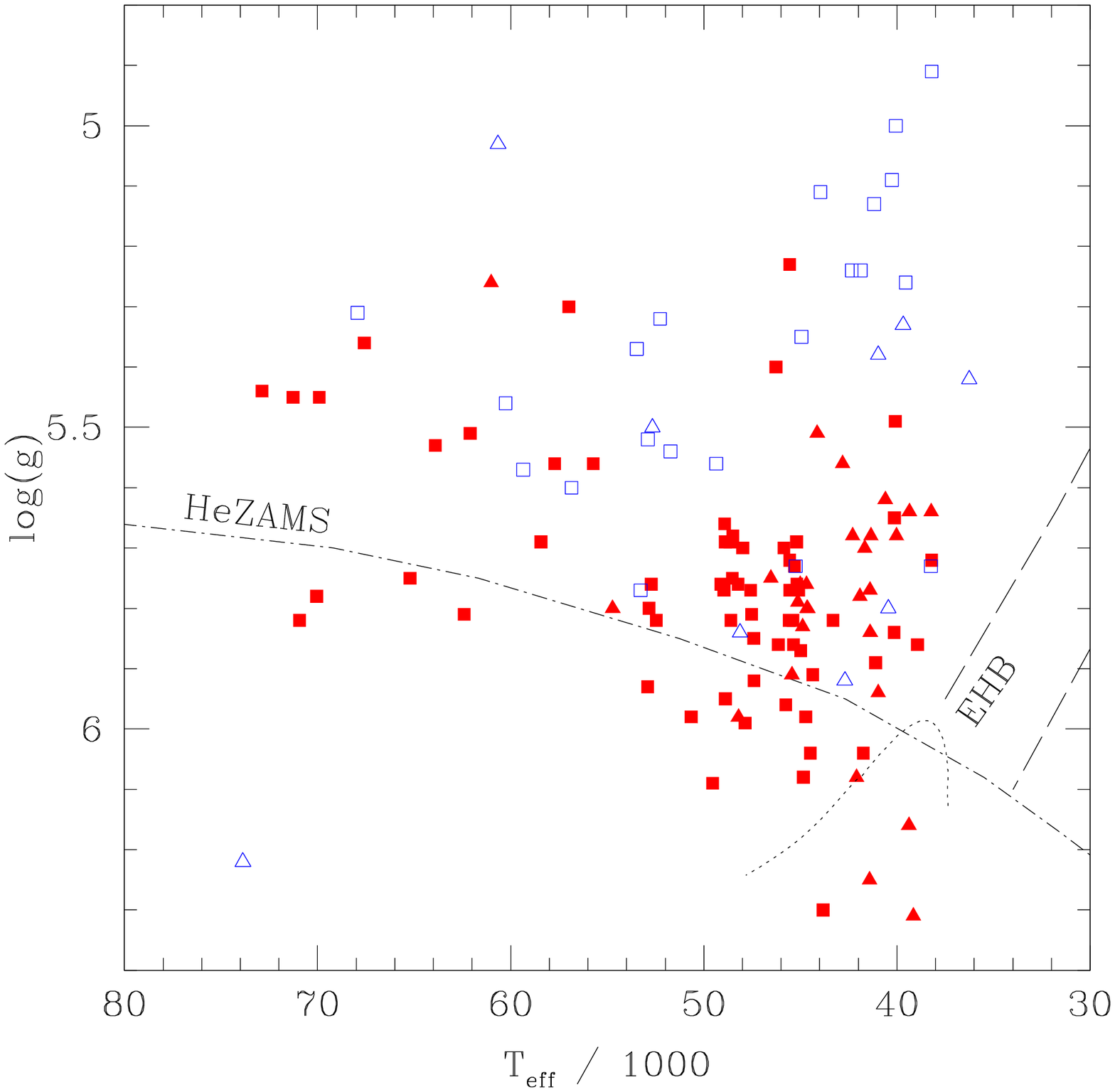}
\caption{{\it Left:} Comparison of our sample to three post-EHB tracks
\citep{dorman93} for $0.475\,M_\mathrm{\odot}$,
$0.473\,M_\mathrm{\odot}$ and $0.471\,M_\mathrm{\odot}$ (top to
bottom).  In the upper left two post AGB tracks \citep{schoen83} for
$0.565\,M_\mathrm{\odot}$ and $0.546\,M_\mathrm{\odot}$ (left to
right) are shown as well.  They may explain a few of the very hot sdO
stars.  {\it Right:} A track of a late hot flasher after it has settled onto
the EHB is plotted \citep{swei97,moehler02}.  Symbols as in
Fig.~\ref{img:hrd}.}
\label{img:evo}
\end{figure}

The canonical evolution theory considers sdO stars to be the progeny
of the sdB stars irrespective of their helium content.  Thus a star
reaching the EHB will evolve from the EHB towards the sdO regime with
higher temperatures and eventually to the white dwarf cooling curve.
Such tracks have been calculated by \citet{dorman93}.  They started
with a star at the tip of the RGB and calculated the minimum core mass
needed to ignite helium burning, then added a small envelope.  The
needed core mass varies slightly and ranges from
$0.464\,M_\mathrm{\odot}$ for supersolar to $0.495\,M_\mathrm{\odot}$
for subsolar metallicity, while the envelope mass is very small
($M_{\mathrm{env}} \le 0.002\,M_\mathrm{\odot}$).  Still, the question
for the cause of the mass loss remains unanswered.  In
Fig.~\ref{img:evo} we compare three post-EHB evolution tracks to the
observed distribution.  They follow the evolution from the zero age
horizontal branch, where core helium burning starts, to the terminal
age horizontal branch, and beyond, for stars with 0.471, 0.473 and
0.475 solar mass cores.  Because of the small envelopes, the post EHB
evolution hydrogen shell burning is insignificant and therefore the
star does not climb the AGB (ABG manqu\'e stars), but evolves through
higher temperatures until eventually it settles on the white dwarf
cooling curve.  These post-EHB tracks cover the sdB regime as well as
the lower gravity helium-deficient sdO stars at $T_\mathrm{eff} \ge
40\,000\,\mathrm{K}$.  This is consistent with the helium-deficient
sdO stars being the descendents of the sdB stars.

However, some of the hottest stars and in particular the
helium-enriched stars which cluster at $T_\mathrm{eff} =
45\,000\,\mathrm{K}$ do not fit into this scheme.  The very hot sdO
stars may be explained as post-AGB stars on their way to the white
dwarf cooling curve.  Two such tracks calculated by \citet{schoen83}
are shown in Fig.~\ref{img:evo}.  But due to the very short
evolutionary timescales ($\sim30\,000\,\mathrm{yrs}$ from ejection of
the planetary nebula to pre-white dwarf settling) for the post-AGB
stars, we expect only a few such stars in our sample.

For the helium-enriched sdOs on the other hand, we have to find
another mechanism that explains both their distribution as well as the
peculiar carbon/nitrogen line strengths in connection with the high
helium abundances.  \citet{swei97} and \citet{brown01} analyzed a
scenario for stars called late hot flasher.  They argue, that fast
rotation of RGB stars could lead to mixing of helium into the
envelope, resulting in a higher RGB peak luminosity and therefore to
higher mass loss rates on the RGB.  This will delay the helium flash
until the star has left the RGB and is already descending on the white dwarf
cooling curve.  The delayed flash will induce mixing which will
transport hydrogen into the helium burning core where it is burnt,
resulting in a helium-enriched star with $T_\mathrm{eff} \approx
40\,000\,\mathrm{K}$ on or near the helium main sequence and enriched
in carbon.

We have plotted an evolutionary track resulting from a late hot flasher in Fig.~\ref{img:evo}.
Also this track fails to reproduce the observed distribution, though the very late hot flasher scenario can explain stars below the helium main sequence.

\section{Carbon Abundances from High-Resolution Spectra (SPY)}
It is clear now, that the helium abundance and the
$T_\mathrm{eff}$--$\log g$ diagram alone cannot answer the question as
to the sdOs' origin.  As different scenarios predict different
abundance patterns in particular for carbon and nitrogen, we started
with a carbon abundance analysis of high resolution spectra of the sdO
stars from SPY.  The new models include C\,{\sc III} and C\,{\sc IV}
in addition to hydrogen and helium.  They were calculated and fitted
to the observed spectrum using the same methods as descibed above.

\subsection{Preliminary Results}
\begin{figure}[ht]\centering
\includegraphics[height=\textwidth,width=7cm,angle=-90,bb=160 30 571 741]{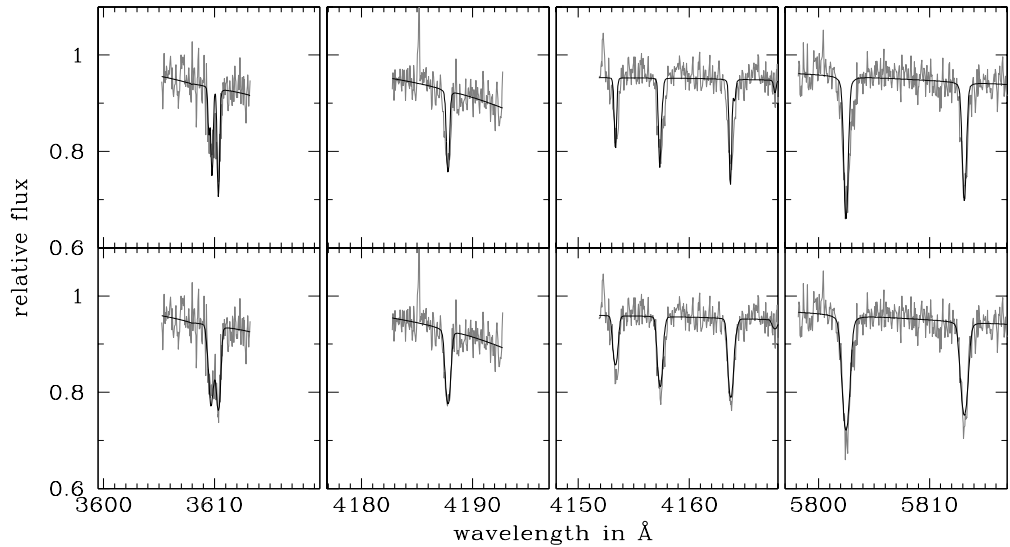}
\caption{Line fits to C\,{\sc III}\,3609.0, 3609.6\,\AA, C\,{\sc
III}\,4186.9\,\AA, C\,{\sc III}\,4152.4, 4156.5, 4162.8\,\AA\ and
C\,{\sc IV}\,5801.3, 5812.0\,\AA\ (from left to right).  Top panels: No
rotational broadening, bottom panel: $v_\mathrm{rot} \sin i =
25\,\mathrm{km\,s^{-1}}$.}
\label{img:c}
\end{figure}

So far we have investigated seven helium-enriched sdOs.
All stars are slightly enriched in carbon, with $\log (\frac{N_\mathrm{C}}{N_\mathrm{total}}) = -2.6 \ldots -3.0$, compared to the solar value of $\log (\frac{N_\mathrm{C}}{N_\mathrm{total}}) = -3.5$.
However, when fitting the carbon lines, two stars showed a significant line broadening indicating projected rotational velocities of $25\,\mathrm{km\,s^{-1}}$ (see Fig.~\ref{img:c}) and $30\,\mathrm{km\,s^{-1}}$, which is surprising as most sdB stars do not show $v_\mathrm{rot} \sin i > 10\,\mathrm{km\,s^{-1}}$ (see Geier et al., these proceedings).
Though low number statistics, this might be considered another hint pointing towards the assumption that sdB stars and helium-enriched sdO stars are two different populations that formed by different evolutionary paths.

\section{Summary and Conclusion}
We have fitted synthetic NLTE model spectra to 87 sdO spectra from the
SDSS database, using a $\chi^2$ technique.  While the helium-enriched
stars cluster in a relative small regime around $T_\mathrm{eff} =
45\,000\,\mathrm{K}$, the helium-deficient sdO stars are widely
spread.  Worth noting is a significant number of helium-deficient sdOs
\emph{below} the helium main sequence, a fact that cannot be explained
by canonical evolution.

Preliminary results of a carbon abundance analysis shows slightly supersolar values for seven helium-enriched sdOs.
The higher $v_\mathrm{rot} \sin i$ we found in two of them compared to the very slowly rotating sdBs might be an additional support for the merger scenario.

By comparing the stars' distribution in the $T_\mathrm{eff}$--$\log
g$ diagram to several evolutionary paths, we conclude that most
likely the helium-deficient sdO stars are the progenies of the sdB
stars.  The helium-enriched sdO stars still pose a problem: The late
hot flasher scenario as well as the white dwarf mergers qualify as
formation channels as they predict an enrichment of helium and
carbon/nitrogen.  The merger scenario is attractive because it can
also naturally explain the absence of radial velocity variables (close
binaries) amongst the helium-enriched sdOs.

\acknowledgements 
We thank Thomas Rauch for providing us with his C\,{\sc III} and
C\,{\sc IV} model atoms and Simon H\"{u}gelmeyer for his introduction
and assistence with the PRO2 code.  This work was supported by DFG
through grant HE1356/44-1.


\begin{thebibliography}{}
\bibitem[Brown et al.(2001)]{brown01}
Brown, T. M., Sweigart, A. V., Lanz, T., Landsman, W. B., \& Hubeny, I. 2001, ApJ, 562, 368
\bibitem[Dorman, Rood \& O'Connell(1993)]{dorman93}
Dorman, B., Rood, R. T., \& O'Connell, R. W. 1993, ApJ, 419, 596
\bibitem[Dreizler(2003)]{dreizler03} Dreizler, S. 2003, in ASP
Conf. Ser. 288, Stellar Atmosphere Modeling, eds.\ I. Hubeny,
D.\ Mihalas, \& K.\ Werner, (San Francisco: ASP), 69
\bibitem[Driebe et al.(1998)]{driebe98}
Driebe, T., Sch\"{o}nberner, D., Bl\"ocker, T., 
\& Herwig, F. 1998, A\&A, 339, 123
\bibitem[Gourgouliatos \& Jeffery(2006)]{gour06}
Gourgouliatos, K. N., \& Jeffery, C. S. 2006, MNRAS, 371, 1381
\bibitem[Han et al.(2002)]{han02}
Han, Z., Podsiadlowski, Ph., Maxted, P. F. L., Marsh, T. R., \& Ivanova, N. 2002, MNRAS, 336, 449
\bibitem[Han et al.(2003)]{han03}
Han, Z., Podsiadlowski, Ph., Maxted, P. F. L., \& Marsh, T. R. 2003, MNRAS, 341, 669
\bibitem[Heber(1986)]{heber86}
Heber, U. 1986, A\&A, 155, 33
\bibitem[Heber et al.(2003)]{heber03}
Heber, U., Edelmann, H., Lisker, T., \& Napiwotzki, R. 2003, A\&A, 411, 477
\bibitem[Heber et al.(2003)]{heber03wind} Heber, U., Maxted, P. F. L.,
Marsh, T. R., Knigge, C., \& Drew, J. E. 2003, in ASP Conf. Ser. 288,
Stellar Atmosphere Modeling, eds.\ I. Hubeny, D.\ Mihalas, \& K.~Werner, 
(San Francisco: ASP), 251
\bibitem[Moehler et al.(2002)]{moehler02}
Moehler, S., Sweigart, A. V., Landsman, W. B., \& Dreizler, S. 2002, A\&A, 395, 37
\bibitem[Napiwotzki(1999)]{napi99}
Napiwotzki, R. 1999, A\&A, 350, 101
\bibitem[Napiwotzki et al.(2001)]{napi01}
Napiwotzki, R., Christlieb, N., Drechsel, H., Hagen, H.-J., Heber, U., Homeier, D., Karl, C., Koester, D., Leibundgut, B., Marsh, T.~R., Moehler, S., Nelemans, G., Pauli, E.-M., Reimers, D., Renzini, A., \& Yungelson, L. 2001, Astronomische Nachrichten, 322, 411
\bibitem[Napiwotzki et al.(2004)]{napi04}
Napiwotzki, R., Karl, C.~A., Lisker, T., Heber, U., Christlieb, N., Reimers, D., Nelemans, G., \& Homeier, D. 2004, Ap\&SS, 291, 321
\bibitem[Saio \& Jeffery(2000)]{saio2000}
Saio, H., \& Jeffery, C. S. 2002, MNRAS, 313, 671
\bibitem[Sch\"{o}nberner(1983)]{schoen83}
Sch\"{o}nberner, D. 1983, ApJ, 272, 708
\bibitem[Str\"{o}er et al.(2007)]{stroer07}
Str\"{o}er, A., Heber, U., Lisker, T., Napiwotzki, R., Dreizler, S., Christlieb, N., \& Reimers, D. 2007, A\&A, 462, 269
\bibitem[Sweigart(1997)]{swei97}
Sweigart, A. V. 1997, in The Third Conference on Faint Blue Stars, eds.\
A. G. D. Philip, J. Liebert, R. Saffer., \& D. S. Hayes, (Schenectady: L. 
Davis Press), 3
\bibitem[Werner \& Dreizler(1999)]{werner99}
Werner, K., \& Dreizler, S. 1999, arXiv:astro-ph/9906130
\end{thebibliography}
\end{document}